\pdfoutput=1 
\documentclass[cits]{JINST}

\title{Prototype of an angular-selective photoelectron calibration source for the KATRIN experiment}

\author{K. Valerius$^a$\thanks{Corresponding author}~\thanks{Present address: Physikalisches Institut, Friedrich-Alexander-Universit\"at Erlangen-N\"urnberg, D-91058 Erlangen, Germany}~, H. Hein$^a$, H. Baumeister$^a$, M. Beck$^a$\thanks{Present address: Institut f\"ur Physik, Johannes Gutenberg-Universit\"at Mainz, D-55099 Mainz, Germany}~, K. Bokeloh$^a$\thanks{n\'{e}e K. Hugenberg}~, J. Bonn$^b$, F. Gl\"uck$^{c,d}$, H.-W. Ortjohann$^a$, B. Ostrick$^{a,b}$, M. Zbo\v{r}il$^{a,e}$ ~and Ch. Weinheimer$^a$\\
\llap{$^a$}Institut f\"ur Kernphysik, Westf\"alische Wilhelms-Universit\"at M\"unster, \\D-48149 M\"unster, Germany\\
\llap{$^b$}Institut f\"ur Physik, Johannes Gutenberg-Universit\"at Mainz, \\D-55099 Mainz, Germany\\
\llap{$^c$}Institut f\"ur Experimentelle Kernphysik, KIT, \\D-76131 Karlsruhe, Germany\\
\llap{$^d$}KFKI, RMKI, \\H-1525 Budapest, POB 49, Hungary\\
\llap{$^e$}Nuclear Physics Institute ASCR,\\CZ-25068 \v{R}e\v{z} near Prague, Czech Republic\\~\\
E-mail: \email{valerius@uni-muenster.de}
}

\abstract{The method of direct neutrino mass determination based on the kinematics of tritium beta decay, which is adopted by the KATRIN experiment, makes use of a large, high-resolution electrostatic spectrometer with magnetic adiabatic collimation. In order to target a sensitivity on $m(\nu)$ of $\unit[0.2]{eV/c^2}$, a detailed understanding of the electromagnetic properties of the electron spectrometer is essential, requiring comprehensive calibration measurements with dedicated electron sources. In this paper we report on a prototype of a photoelectron source providing a narrow energy spread and angular selectivity. Both are key properties for the characterisation of the spectrometer. The angular selectivity is achieved by applying non-parallel strong electric and magnetic fields: Directly after being created, photoelectrons are accelerated rapidly and non-adiabatically by a strong electric field before adiabatic magnetic guiding takes over.}


\usepackage{units}
\usepackage{amsmath}

\begin{document}

\newcommand{\K}{KATRIN}
\newcommand{\KE}{KATRIN experiment}
\newcommand{\KM}{KATRIN main spectrometer}
\newcommand{\KP}{KATRIN pre-spectrometer}
\newcommand{\KC}{KATRIN collaboration}
\newcommand{\Mac}{MAC-E filter}
\newcommand{\Sim}{Sim\-ion}
\newcommand{\Bfield}{Bfield\_3d}
\newcommand{\ie}{i.e.}
\newcommand{\eg}{e.g.}
\newcommand{\etal}{et al.}

\newcommand{\entspricht}{\mbox{$\stackrel{^}{=}$}}
\newcommand{\sollsein}{\mbox{$\stackrel{\mathrm{!}}{=}$}}
\newcommand{\degree}{\ensuremath{^{\circ}{}}}  	
\newcommand{\belec}{$\beta$ electron}

\section{Introduction}
\label{sec:introduction}
Electrostatic spectrometers with magnetic adiabatic guiding (so-called {\Mac}s) have been used in the past to scan the endpoint region of the tritium $\beta$-spectrum for the tiny signature of the neutrino mass. Based on this direct kinematic method, neutrino mass searches led by groups in Mainz and Troitsk have established an upper limit of $m(\nu_\mathrm{e}) < \unit[2]{eV/c^2}$ \cite{kraus-paper,lobashev-upperlimit}. The \KE\ \cite{kdr,proceeding-beck} is currently being set up at Karlsruhe with the aim of improving the sensitivity of the method by an order of magnitude, \ie\ down to the level of $\unit[0.2]{eV/c^2}$. Such an improvement implies -- among other requirements -- an energy resolution of the spectrometer of $\Delta E \lesssim \unit[1]{eV}$ for $\unit[18.6]{keV}$ electrons. From this aim the need to provide electron sources capable of probing the characteristics of such a high-resolution electron spectrometer with very good precision arises. 
In particular, such a calibration source should exhibit an intrinsic energy spread smaller than the design resolution of the spectrometer as well as the capability to select both the transversal position within the magnetic flux tube and the emission angle with respect to the magnetic field direction.
The latter requirement is derived from the fact that the energy resolution of a \Mac\ corresponds directly to the distribution of the starting angles of the particles under investigation (see Figure~\ref{fig:transfkt-mainspec} in Section \ref{sec:motivation}).

This article is organised as follows: 
After briefly reviewing the \Mac\ technique with special emphasis on the requirements for a suitable calibration electron source in Section \ref{sec:motivation}, we illustrate the underlying principles and the conceptual design of such a source in Section \ref{sec:experimental-principle}. Experimental results of test measurements carried out with our prototype source will be described in Section \ref{sec:measurements}. In Section \ref{sec:discussion} we discuss our results and give an outlook on potential future developments.

\section{Motivation: an angular-resolved calibration source for KATRIN}
\label{sec:motivation}
\subsection{The {\Mac} technique}
\label{subsec:mac-e-technique}
The principle of electrostatic filtering combined with magnetic adiabatic collimation (\eg\ \cite{picard-nimb,lobashev85}) 
was developed in order to overcome a basic disadvantage of previous high-resolution magnetic spectrometers, namely their low angular acceptance and hence limited accepted luminosity. Since the {\Mac} is an integrating spectrometer, it is ideally suited for low count-rate spectrometry measurements at the upper end of the energy spectrum of charged particles, like kinematic neutrino mass searches using tritium $\beta$-decay \cite{otten-weinheimer-review} or precision weak interaction studies \cite{witch,aSpect}. Although {\Mac}s are being used for spectroscopy of electrons as well as of positively charged ions, in the following we will explain its principle assuming an application in electron spectroscopy like in the \KE, for which our electron source can be used.

The experimental set-up of a \Mac\ comprises a vacuum vessel containing a high-voltage electrode system that provides the electrostatic retardation potential, and a chain of magnets to produce the magnetic field which guides the electrons on cyclotron trajectories from the source through the spectrometer and eventually on to the detector. While at any point throughout the set-up the magnetic field strength must be high enough to maintain an adiabatic motion of the electrons, its absolute value varies by several orders of magnitude in order to achieve the magnetic collimation effect, as will be explained below.

The total kinetic energy $E_\mathrm{kin}$ of the electrons is split into two contributions: one associated with the motion parallel to the magnetic field lines ($E_\parallel$), and the second one in the transverse direction (cyclotron component, $E_\perp$):
\begin{equation}
\begin{array}{lcl}
E_\parallel 	
	& = & E_\mathrm{kin} \cdot \cos^2\theta,\\[2mm]
E_\perp	& = & E_\mathrm{kin} \cdot \sin^2\theta\\[1mm]
	& = & E_\mathrm{kin} - E_\parallel ,
\label{equ:eperp}
\end{array}
\end{equation}
where $\theta$ denotes the angle subtended by the electron momentum $\vec{p}$ and the local magnetic field $\vec{B}$.

The cyclotron motion gives rise to an orbital magnetic moment $\mu$, which in the adiabatic approximation\footnote{The motion of the electron along the magnetic field line is considered to be adiabatic as long as the relative change in the magnetic field strength per cyclotron turn remains sufficiently small. This condition can be expressed in terms of the cyclotron frequency $\omega_\mathrm{cyc}$ as $\left| \frac{1}{B}\,\frac{\mathrm{d}\vec{B}}{\mathrm{d}t} \right| \ll \omega_\mathrm{cyc}$.} defines a constant of the motion:
\begin{equation}
\mu = \frac{E_\perp}{B} = \mathrm{const.}
 \label{equ:adiabinv-nonrelativistic}
\end{equation}
Here, we employ the non-relativistic limit.\footnote{In a relativistic treatment the conserved quantity is given by $\gamma\mu$, where $\gamma$ denotes the relativistic Lorentz factor.} 
From Eqs.~(\ref{equ:eperp}) and (\ref{equ:adiabinv-nonrelativistic}) one derives the following transformation law for the angle $\theta$ for the case without electrical retarding or accelerating fields (\ie, $E_\mathrm{kin} = \mathrm{const.}$):
\begin{equation}
\frac{\sin^2\theta_f}{\sin^2\theta_i} = \frac{B_f}{B_i}.
\label{equ:angle-transform}
\end{equation}
This expression states that the initial angle $\theta_i$ of an electron starting in a strong magnetic field $B_i$ will be gradually lowered to a value $\theta_f < \theta_i$ as it moves into a weaker magnetic field $B_f < B_i$, or, conversely, steepened for an electron travelling from a low into a high magnetic field.\footnote{The latter scenario is of course well known as the `magnetic mirror' effect, \eg\ used to form `magnetic bottles'.} Equation (\ref{equ:adiabinv-nonrelativistic}) also holds in case of non-zero electric fields in the adiabatic limit, since the gain or loss of transversal energy by an electric field will be averaged to zero by the cyclotron motion. 

By choosing the magnetic field $B_\mathrm{min} = B_\mathrm{ana}$ at the central analysing plane of the spectrometer 
(where the electrostatic retardation potential reaches its maximum) much weaker than the maximum magnetic field strength $B_\mathrm{max}$ occurring in the set-up, a minimisation of the residual and non-analysable cyclotron energy component $E_\perp$ in the analysing plane can be achieved. Hence, the energy resolution $\Delta E$ of a {\Mac} at a given electron energy $E_0$ is solely determined by the ratio of $B_\mathrm{min}$ and $B_\mathrm{max}$:
\begin{equation}
\Delta E = E_0 \cdot
\frac{B_\mathrm{min}}{B_\mathrm{max}}.
\label{equ:energy-resolution}
\end{equation}
Apart from its general relevance for the {\Mac} technique, Eq.~(\ref{equ:angle-transform}) is also a key to the working principle of the
angular-selective electron gun we present in this paper (see Section \ref{sec:experimental-principle}).

\subsection{Transmission properties of the \Mac}
\label{subsec:transmission}
\begin{figure}[!htb]
 \centering
\includegraphics[angle=-90,width=0.7\textwidth]{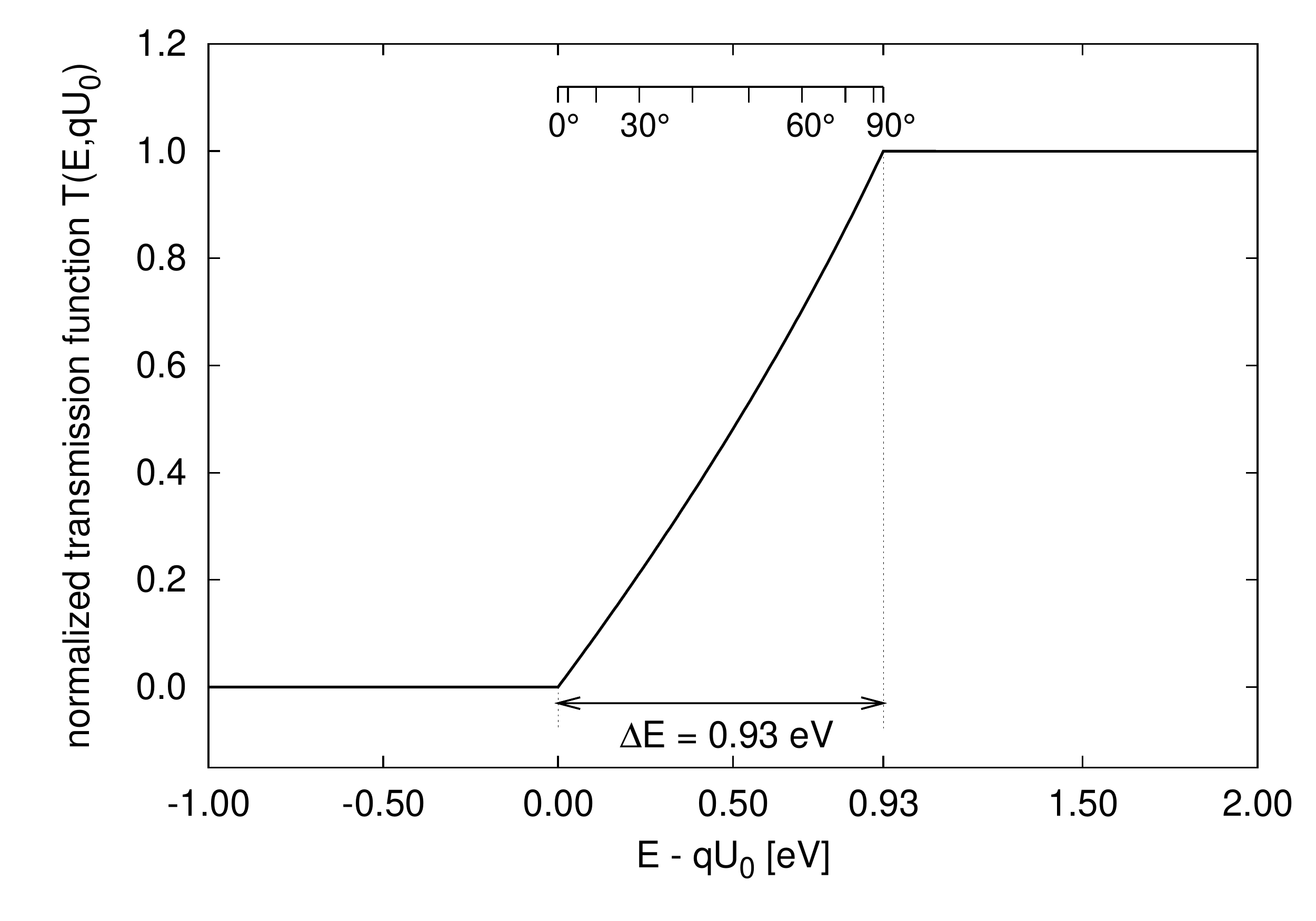}
\caption{Transmission function of the KATRIN main spectrometer
for an isotropically emitting electron source. The proposed magnetic field settings $B_\mathrm{source} = \unit[3.6]{T}$, $B_\mathrm{max} = \unit[6]{T}$ and $B_\mathrm{min} = \unit[0.3]{mT}$ result in an energy resolution of $\Delta E = \unit[0.93]{eV}$ at $E = \unit[18.6]{keV}$. 
The angular scale (at $B=B_\mathrm{max}$) illustrates that electrons with larger angles are transmitted only at higher 
surplus energies $E-qU_0$ because of their larger cyclotron energy component. The feature of magnetic adiabatic collimation in principle allows to collect electrons with angles up to almost $\unit[90]{^\circ}$ in the maximum magnetic field.} \label{fig:transfkt-mainspec}
\end{figure}

The transmission function of an ideal {\Mac} can be expressed analytically \cite{picard-nimb} as a function of the electric retardation potential $U_0$ of the filter and of the magnetic field strengths at the electron source ($B_\mathrm{source}$), at the analysing plane ($B_\mathrm{ana}$) and at the pinch magnet ($B_\mathrm{max}$). Its slope and width for an isotropically emitting
electron source
are defined only by the magnetic field ratios $B_\mathrm{ana}/B_\mathrm{source}$ and $B_\mathrm{ana}/B_\mathrm{max}$, respectively. Figure~\ref{fig:transfkt-mainspec} shows the ideal transmission function computed for the electric potential and magnetic field settings of the {\KM}. 

However, for a realistic set-up the well defined and sharp edges of the transmission function get smeared out by radial inhomogeneities of the electrostatic potential, $\Delta U_\mathrm{ana}$, and of the magnetic field strength, $\Delta B_\mathrm{ana}$, across the analysing plane. Furthermore, the combined effect of both imperfections leads to a broadening of the transmission function, which is equivalent to a deterioration of the energy resolution $\Delta E$. Typical values of the field inhomogeneities which can be achieved at the {\KM} are of the order $\Delta U_\mathrm{ana} \approx 1\;\mathrm{V}$ and $\Delta B_\mathrm{ana} \approx 0.03\;\mathrm{mT}$ across the radius of $r_\mathrm{ana} = 4.5\;\mathrm{m}$ of its analysing plane.\footnote{Hence, the relative inhomogeneities amount to $\Delta U_\mathrm{ana}/U_\mathrm{ana} \approx 5 \cdot 10^{-5}$ and $\Delta B_\mathrm{ana}/B_\mathrm{ana} \approx 10\%$.}  The broadening of the transmission function would be significant, and the energy resolution may be worsened to about 2 -- 3 times its nominal value of $\Delta E = 0.93\;\mathrm{eV}$ if one would average over the full magnetic flux tube. In order to avoid this large broadening, the KATRIN experiment uses a 13-fold radially (and 12-fold azimuthally) segmented detector. Thus the broadening of the transmission function for each individual detector pixel is reduced by about a factor 13.

The KATRIN main spectrometer uses a system of double-layer wire electrodes \cite{kdr,valerius_erice} to avoid the background from secondary electrons emitted from the spectrometer walls by cosmic muons or by environmental radioactivity. The electrons on the outermost trajectories, especially those with large starting angles with respect to the magnetic field direction, are very sensitive to misalignments or failures of any of the 248 electrode modules. The time of flight of such large-angle electrons on outer radii can even help to localise the origin of a potential problem regarding the electrode system\footnote{The \Mac\ can be operated in a so-called MAC-E-TOF (time of flight) mode \cite{bonn99,valerius09b}, which transforms the spectrometer into a non-integrating one. The principle of the MAC-E-TOF mode can be also used for studying the correctness of the electric retarding potential along the trajectories by investigating the time-of-flight of the electrons.} \cite{valerius09b}. Of course these special investigations require well-defined electron energies.

\subsection{Design considerations for a calibration electron source}
The previous description of the spectrometer characteristics leads to the following requirements for a calibration source:
\label{subsec:designcriteria}
\begin{enumerate}
 \item The intrinsic energy spread $\delta E$ of the produced electrons should be small (in particular: smaller than the energy resolution $\Delta E_\mathrm{spec}$ of the spectrometer), limiting the allowed relative energy broadening $\delta E/E$ to values of the order of $10^{-5}$.
 \item The electron beam intensity should be tunable, depending on the specific calibration purpose at hand, ranging from a few 100 electrons per second
to several $10^4$ electrons per second.
 \item Pulsed operation with pulse lengths on the order of some ns up to several 10 $\mu$s should be possible to enable time-of-flight (TOF) studies.
 \item The emission of electrons should be spatially very well confined (point-like), and the spot should be movable across the full extension of the magnetic flux tube to test the transmission condition for individual pixels of the multi-pixel detector.
 \item Good angular selectivity should be achieved to put special emphasis on large-angle electrons passing along the outermost field lines of the magnetic flux 
  tube.
\end{enumerate}

\section{Method and experimental set-up}
\label{sec:experimental-principle}
How can the requirements listed in Section \ref{subsec:designcriteria} be technically realised in the concept of a calibration source for the \KE? 
The first three criteria on the list can be fulfilled by using a narrow-band UV LED to produce photoelectrons by irradiating a clean metal surface, as demonstrated in Ref.~\cite{valerius09a}. In particular, the work function $\Phi$ of the metal and the wavelength $\lambda$ of the UV light can be chosen such that a very narrow energy spread of the emitted photoelectrons is achieved. In the earlier work cited above a residual energy spread of $\delta E \approx 0.20$ -- $\unit[0.25]{eV}$ was found for a combination of a stainless steel plate and UV light with $\lambda \approx \unit[265]{nm}$, which is acceptable for our application as a calibration source for \K. The residual spread can for example result from the non-monochromatic emission of the UV LED, from inhomogeneities of the work function across the metal surface, and from fluctuations of the high voltage supplies. The other properties, pulsed emission and beam intensity of the photoelectrons, can be readily accommodated by pulsing the UV LED and varying its driving current. (In Ref.~\cite{valerius09a} pulse durations between $\sim\unit[40]{ns}$ and $\unit[40]{\mu s}$ and switching frequencies up to $\unit[10]{kHz}$
were successfully tested.)

The last two criteria on the list of requirements can be realised by two additional steps: (a) reducing the UV illumination to a small 
spot on the metal surface instead of using a wide-angle UV light beam, and (b) placing the photoelectron source in combined 
inhomogeneous electro- and magnetostatic fields.  These two additional steps, which are new compared to the electron source presented 
in \cite{valerius09a}, will be described in detail in the following sections.

\subsection{Principle of angular-selective photoelectron production}
\begin{figure}[!tb]
\centering
\includegraphics[width=0.45\textwidth]{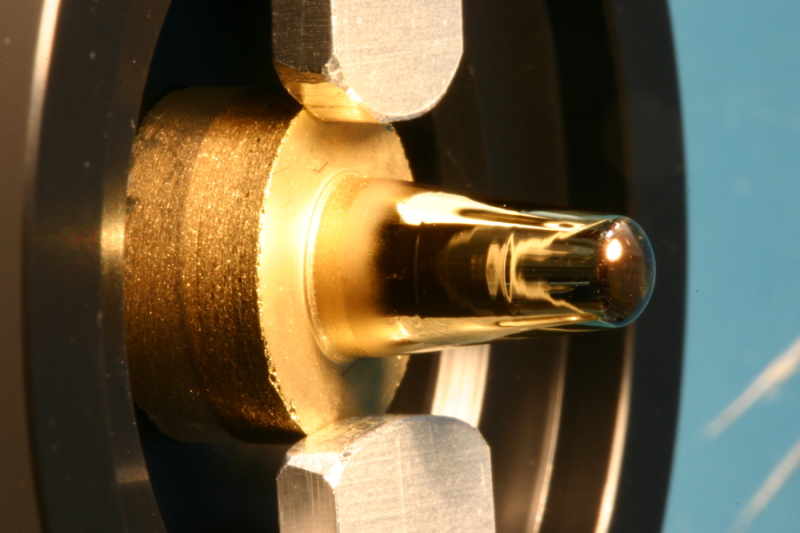}
\caption{Photograph of the gold-plated quartz tip of the photoelectron gun used at the \KP.\label{fig:prespec-egun-foto}}
\end{figure}

\begin{figure}[!tb]
\centering
\includegraphics[width=0.7\textwidth]{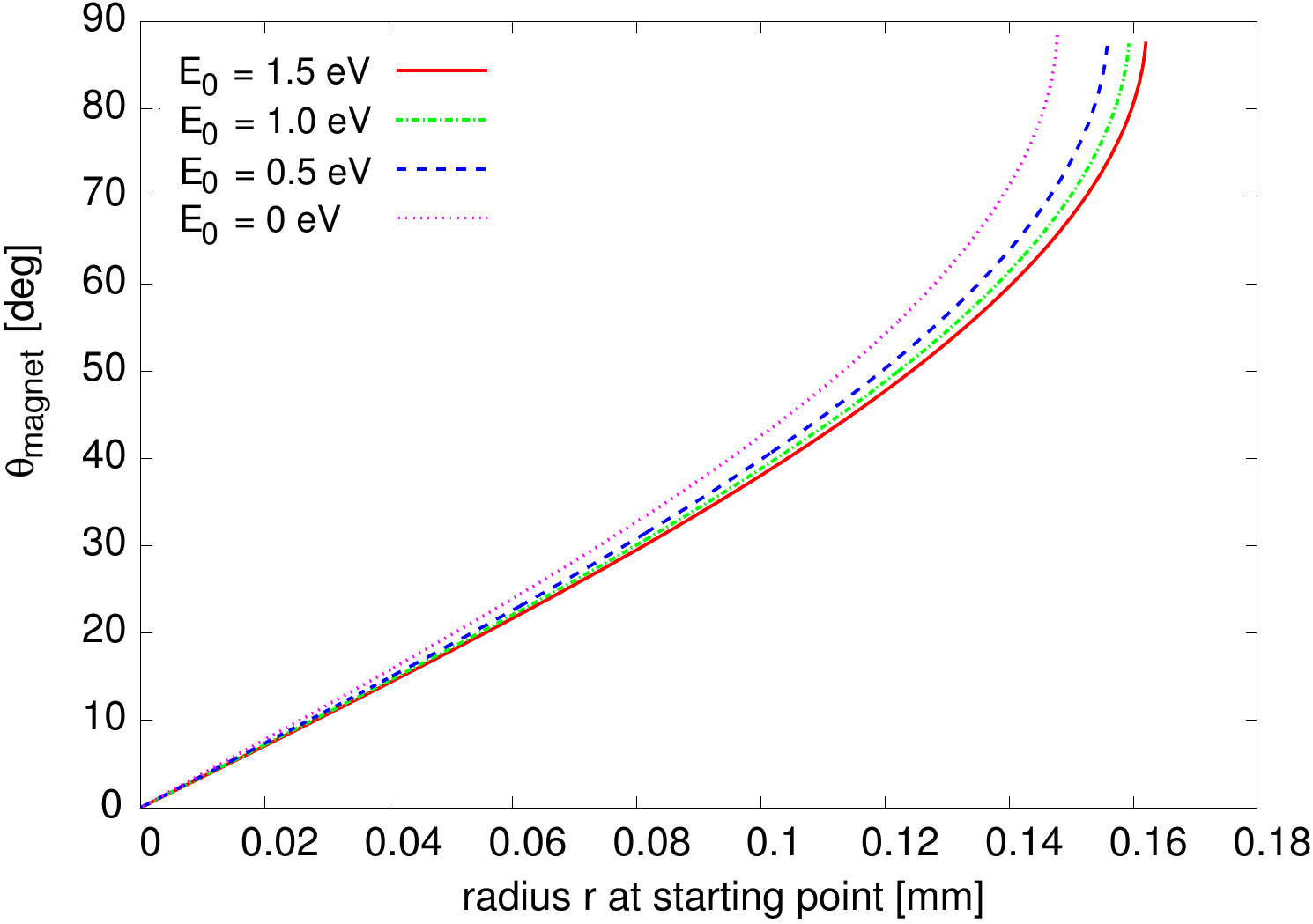}
\caption{Angular emission characteristics of the pre-spectrometer electron gun as a function of the starting radius on the tip \cite{hugenberg08}. 
Simulated values of the angle in the spectrometer solenoid at the entrance $\theta_\mathrm{magnet}$ 
are shown for four discrete values of the starting energy $E_0$ between 0 and $\unit[1.5]{eV}$. \label{fig:prespec-egun-angles}}
\end{figure}

The principle of angular-selective photoelectron production employed in this work was studied during an investigation of the detailed emission characteristics of the electron source used for testing purposes at the \KP\ \cite{hugenberg08}. The pre-spectrometer electron source is similar to the type of source developed for calibration measurements at the Troitsk neutrino mass experiment (see the description in \cite{aseev2000}). The photoelectron source consists of a rounded quartz tip (diameter $\sim\unit[2]{mm}$) coated with a thin layer of gold which is placed on high voltage and irradiated from the rear side by a conventional UV light source. A photograph of the tip is shown in Figure~\ref{fig:prespec-egun-foto}. 
The electron gun is usually installed right outside a strong magnet placed at the entrance of the pre-spectrometer, and hence the electrons will feel a considerable increase in the magnetic field strength (by roughly a factor of 100) when moving in axial direction from the source towards the entrance of the spectrometer. Computer simulations (using the field and tracking codes written by one of the authors \cite{glueck-algorithms}) revealed that most of the photoelectrons are reflected due to the magnetic mirror effect before they can even enter the pre-spectrometer, \ie\ their momentum vector reaches an angle of $\theta = 90^\circ$ somewhere between the source and the centre of the strong magnet placed at the entrance of the spectrometer. This behaviour can be easily understood and quantified with the help of Eq.~(\ref{equ:angle-transform}): assuming adiabatic motion, any small angle $\theta_\mathrm{e-gun} > 0^\circ$ of the electron shortly after leaving the tip in the comparatively low magnetic field $B_\mathrm{e-gun}$ will be increased to
\begin{equation} \label{eq:theta_magnet}
 \theta_\mathrm{magnet} = \arcsin{\left( \sqrt{\frac{B_\mathrm{magnet}}{B_\mathrm{e-gun}}} \cdot \sin \theta_\mathrm{e-gun} \right)}
\end{equation}
as the electron enters the stronger magnetic field $B_\mathrm{magnet}$ inside the bore of the spectrometer solenoid. The point that remains to be clarified is the origin of the starting angle $\theta_\mathrm{e-gun}$. In the absence of an electric field, $\theta_\mathrm{e-gun}$ will be determined by the angular emission profile at the surface of the tip. However, microscopic tracking simulations (see \cite{hugenberg08}) have shown that for a suitable configuration of electric and magnetic fields the angular emission profile at the surface becomes irrelevant (due to the small starting energy of the photoelectrons). Instead, the location of the emission spot on the curved surface of the tip -- more precisely: the radial distance of the point of photoemission from the symmetry axis and, hence, the direction of the local electric field -- defines the angle $\theta$ and thus the relative amount of transversal kinetic energy at the start of the trajectory. This is illustrated in Figure~\ref{fig:prespec-egun-angles}, where the angle $\theta_\mathrm{magnet}$ reached at the center of the pre-spectrometer entrance solenoid is plotted as a function of the radial starting position on the tip. In the simulation, the starting kinetic energy of the electrons was varied between 0 and $\unit[1.5]{eV}$ to take into account the rather broad spectral distribution of UV photons which liberate the photoelectrons, but its influence is marginal.\footnote{This large spectral width is due to the broad-band UV light source employed in the electron gun for the KATRIN pre-spectrometer. For the prototype described in this work, a UV LED with a much narrower spectroscopic range ($\Delta \lambda \approx 35~\mathrm{nm}$) was used.}

\begin{figure}[!tb]
\centering
\includegraphics[width=0.7\textwidth]{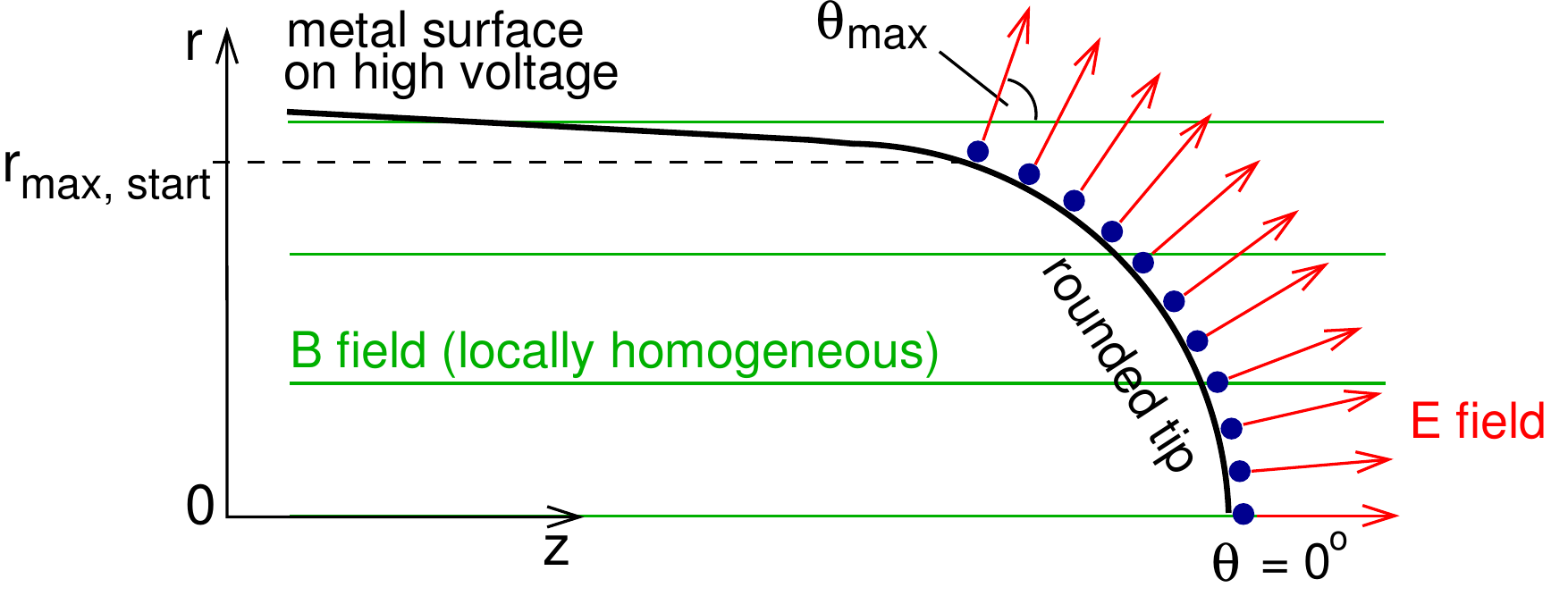}
\caption{Schematic illustration of the relation between electron starting angle and radial position on the tip of the electron gun.\label{fig:fields-eguntip}}
\end{figure}

In turn, we can take advantage of this effect to build an angular-selective photoelectron source. The salient points are that the electric and magnetic fields at the location of the tip must be non-parallel, and that the strength of the magnetic field $B_\mathrm{e-gun}$ at the tip must be low compared to the maximum magnetic field $B_\mathrm{magnet}$. The third prerequisite is that a strong electrostatic acceleration of the electrons should take place at the start of the trajectory, which can be realised by a round photocathode tip on high voltage to achieve a suitable shape and strength of the electric field. In this case, the motion will be non-adiabatic in the initial phase of the motion, and therefore the transformation according to Eq.~(\ref{equ:angle-transform}) will not be valid at the start of the trajectory. As sketched in Figure~\ref{fig:fields-eguntip}, in such a configuration the electrons gain a certain amount of transversal kinetic energy by non-adiabatic acceleration which depends on the angle between electric and magnetic fields. In the early phase of the motion, the electrons follow the electric field lines rather than the magnetic field lines. After a very short distance the electrons leave the region of strongest acceleration. As they have obtained a non-zero velocity $v$, the cyclotron force
\begin{displaymath}
  \vec{F} = -e \vec{v} \times \vec{B}
\end{displaymath} then starts to dominate and averages out any further gain of transversal energy. Subsequently, 
the electrons are guided adiabatically along the magnetic field lines. Due to the specific geometry of the fields, those electrons starting at a larger radial distance from the center of the tip will ``attach`` to the magnetic field lines with a larger angle $\measuredangle(\vec{p},\vec{B})$ than those emitted close to the axis, where $\vec{E}$ and $\vec{B}$ are essentially parallel (compare Fig.~\ref{fig:fields-eguntip}). As they travel from the low magnetic field into a region with higher magnetic field strength, the initial amount of transversal kinetic energy increases, again according to Eq.~(\ref{equ:angle-transform}). In view of a large ratio $B_\mathrm{magnet}/B_\mathrm{e-gun}$, even very small transversal starting energies -- and hence small starting radii $r_\mathrm{start}$ -- are sufficient to yield large angles $\theta_\mathrm{magnet}$ at the center of the entrance solenoid of the spectrometer.

\subsection{Possibilities to realise a limited UV irradiation spot size}
\begin{figure}[!htb]
 \centering\includegraphics[width=0.40\textwidth]{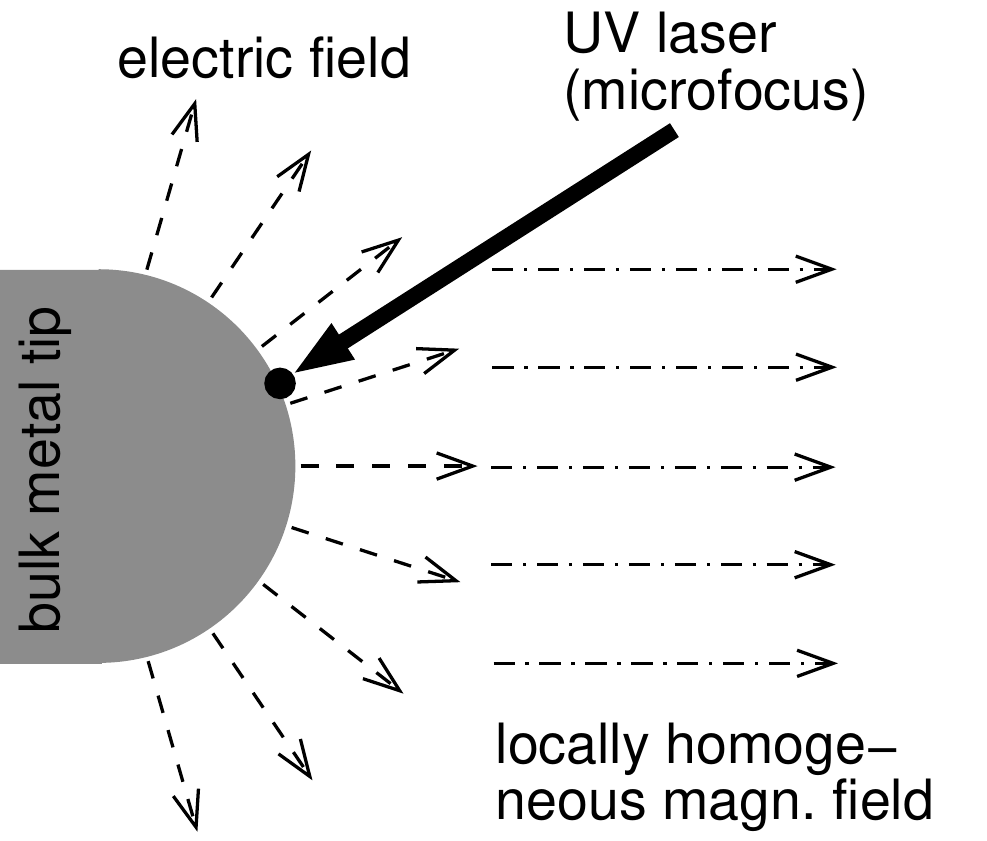}
\hspace{10mm}
\includegraphics[width=0.40\textwidth]{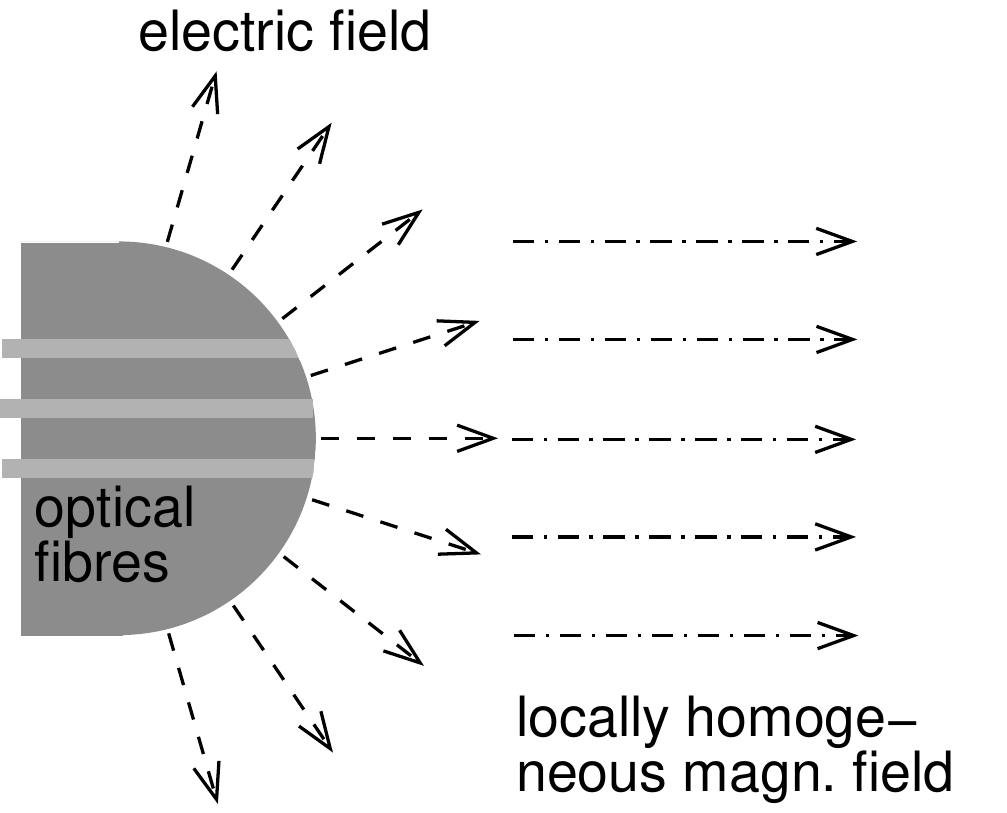}
\caption[Two options of illuminating the photocathode in an angular-selective electron gun]{Two options of irradiating the photocathode in an angular-selective electron gun with a ball-shaped metal tip. {\it Left:} Front-side (reflective) irradiation of a bulk-metal tip by a UV laser. {\it Right:} Rear-side (transmissive) irradiation by UV-transparent optical fibres with metal-coated tips.\label{fig:illumination-options}}
\end{figure}

Based on the discussion of the physical concepts in the previous section, the next problem clearly consists in finding a technique to produce a small emission spot of photoelectrons off a rounded (ideally: ball-shaped) metal surface. The illumination of small, well-defined spots on the surface of a round metal tip may be realised in various ways, two of which are sketched in Fig.~\ref{fig:illumination-options}:
\begin{enumerate}
 \item[(a)] Reflective irradiation -- a small spot on the surface of the bulk metal is irradiated from the front side by a well-collimated beam, for example using a micro-focused UV laser.
 \item[(b)] Transmissive irradiation -- UV-transparent optical fibres are employed to guide the light through the bulk-metal electrode to a position of well-defined radius $r$ on the tip. The fibres are cut and polished at the surface of the tip. Their ends are coated with a thin metal film which is irradiated from the rear side.
\end{enumerate}
Each technique has its advantages and drawbacks (cp. discussion in Ref.~\cite{valerius09b}). We opted for the second method, as it avoids potential problems introduced by reflections of UV light inside the set-up, and because the fixed position of the fibres offers a better reproducibility of the measurements under the given experimental conditions. In this method, the diameter of the light-guiding core of the optical fibre and the minimal possible spacing between the fibres determine the size and position of the irradiated spot, and thus the angular range of photoelectrons expected from each fibre position.

\subsection{Set-up of the fibre-coupled photoelectron gun}
\begin{figure}[!t]
\centering
\includegraphics[width=0.86\textwidth]{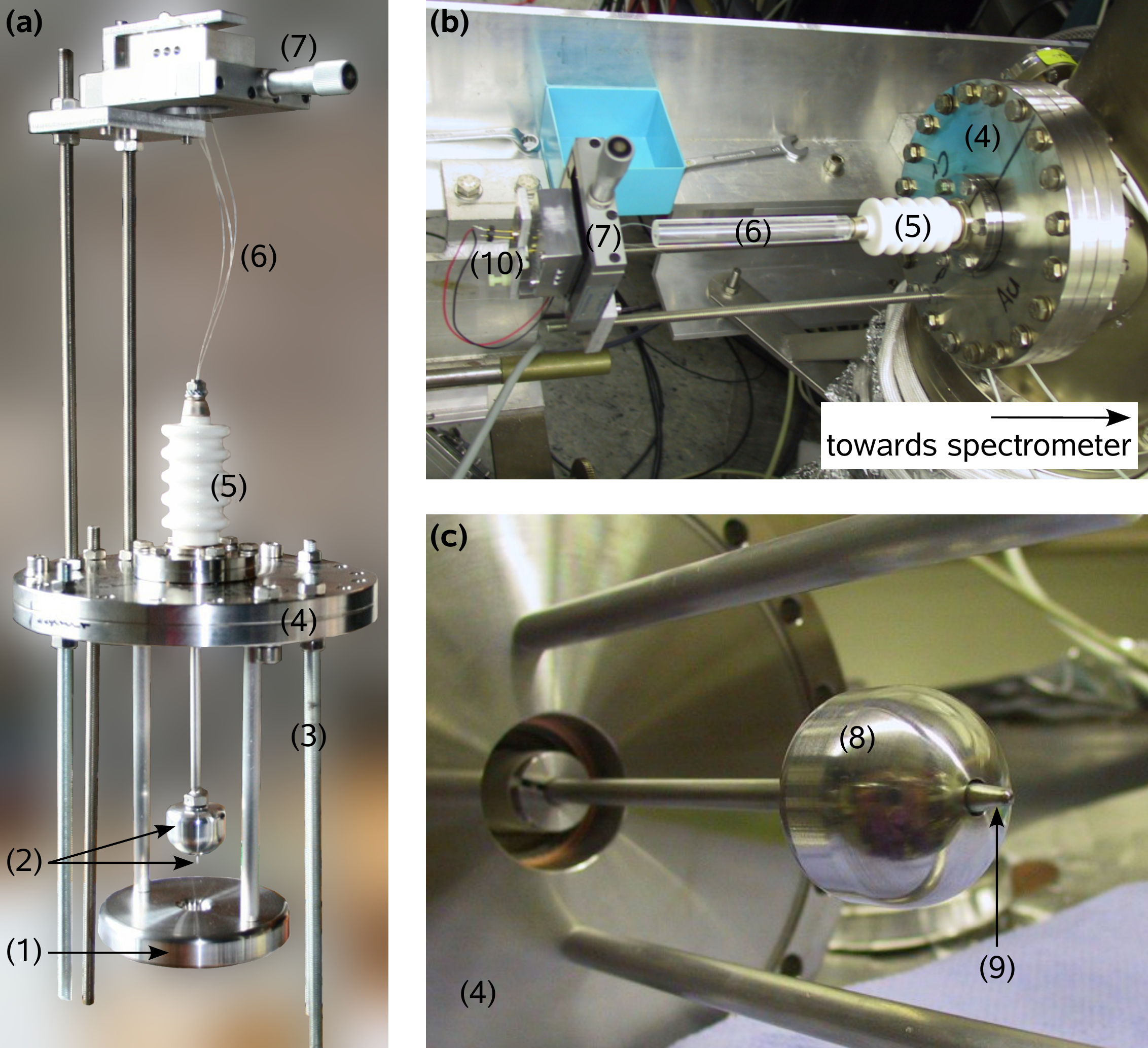}
\caption{Photographs of the prototype photoelectron gun: (a) full set-up, (b) installation at the Mainz \Mac, (c) front/side close-up view. The numbers indicate the following components: (1)~ground electrode, (2)~high voltage electrodes, (3)~support rods, (4)~150\,CF base flange, (5)~high voltage insulator, (6)~fibres, (7)~positioning table to select illuminated fibre, (8)~field-shaping electrode, (9) tip, and (10) UV LED.\label{fig:egun-photos}}
\end{figure}

\begin{figure}[!htb]
 \centering
\includegraphics[width=0.38\textwidth]{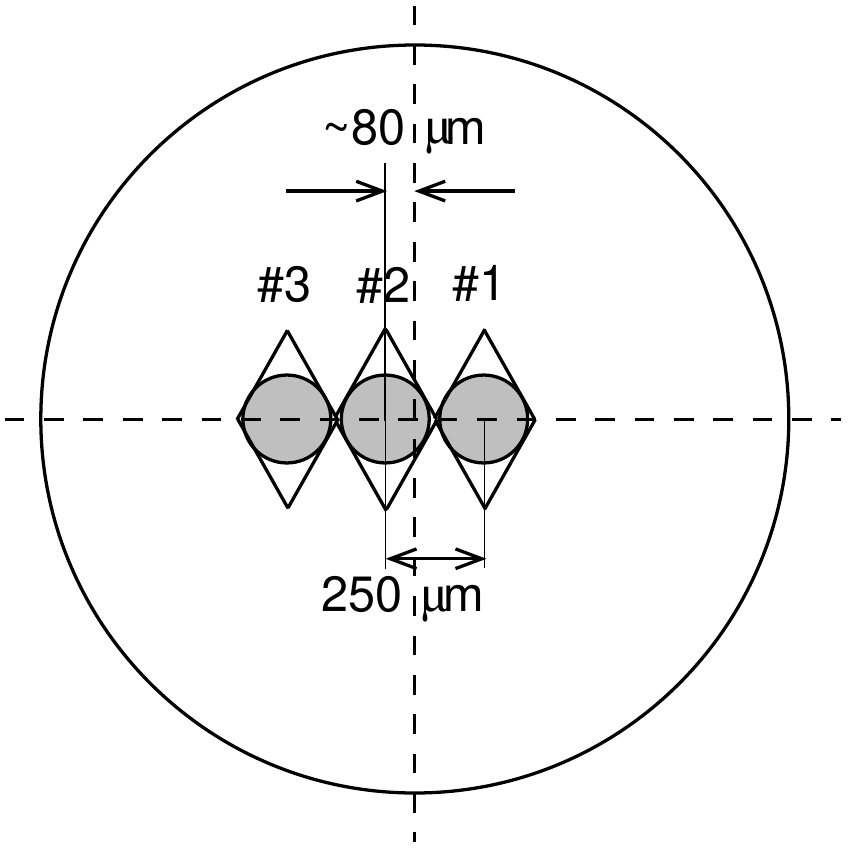}
\caption{Sketch: tip of the electron gun. The cross section of the tip is shown, including the circular outlines of the three fibres implanted closely spaced into the bulk of the tip. The fibres are glued into diamond-shaped openings. The center of fibre \#\,2 is offset by $\unit[(80 \pm 20)]{\mu m}$ with respect to the center of the tip, and fibres \#\,1 and \#\,3 are placed symmetrically around fibre \#\,2. During tests of the set-up before shipment to Mainz, fibre \#\,3 was broken and
could no longer be used. In the description of the measurements below, fibres \#\,1 and \#\,2 will be labelled `outer fibre' and `inner fibre', respectively.\label{fig:drawing-fiberholes}}
\end{figure}

After having laid out the physical principles and technical challenges, the experimental set-up of our prototype electron source shall be described. Figure \ref{fig:egun-photos} presents details of the set-up and of its mounting at the Mainz \Mac. It comprises the following components:
\begin{itemize}
 \item A photocathode tip made of aluminium. To connect the tip with the flange of the vacuum chamber its support structure is equipped with a 30 kV insulator and a high-voltage feedthrough.
 \item A hemispherical casing and an annular ground electrode provide additional shaping of the electric field.
 \item Three UV-transparent fibres are embedded into the metal tip. Step-index UV fibres with a core diameter of $\unit[98]{\mu m}$ and an outer diameter of $\unit[245]{\mu m}$ were chosen. The ends of the fibres were coated with a thin metal film of polycrystalline copper ($\Phi_\mathrm{Cu,\, polycryst.} = \unit[(4.65 \pm 0.05)]{eV}$, see \cite{eastman}), applied in well-defined quantities by the vaporization method.
Figure~\ref{fig:drawing-fiberholes} schematically depicts the arrangement of the three fibres implanted into the metal tip. The minimal spacing is defined by the diameter of the fibre coating, which amounts to $250\;\mathrm{\mu m}$. Note that only the fibres marked as \#\,1 and \#\,2 in Fig.~\ref{fig:drawing-fiberholes} could be used for the prototype test measurements, as the third one was accidentally broken.
%
%
 \item A UV LED was mounted on a positioning table (micrometer screw) to allow the selective illumination of specific fibres. LEDs of the types T9B26C (central wavelength $\lambda_\mathrm{central} = \unit[265]{nm}$) and T9B25C ($\lambda_\mathrm{central} = \unit[255]{nm}$) were employed \cite{seoul-led}. In order to facilitate the interfacing of the LED emission with the optical fibre, these LEDs are equipped with a simple ball lens. Nevertheless, the manual coupling of the UV light beam into the fibre is rather difficult and will lead to considerable losses of UV light intensity. This situation can be improved by using
dedicated LED fibre couplings. We did not use such a coupling, since for our purpose of a first proof-of-principle experiment the photoelectron intensity achieved with the simple method was sufficient.

\end{itemize} 

\section{Experimental Results}
\label{sec:measurements}
\begin{figure}[!htb]
 \centering
\includegraphics[width=0.9\textwidth]{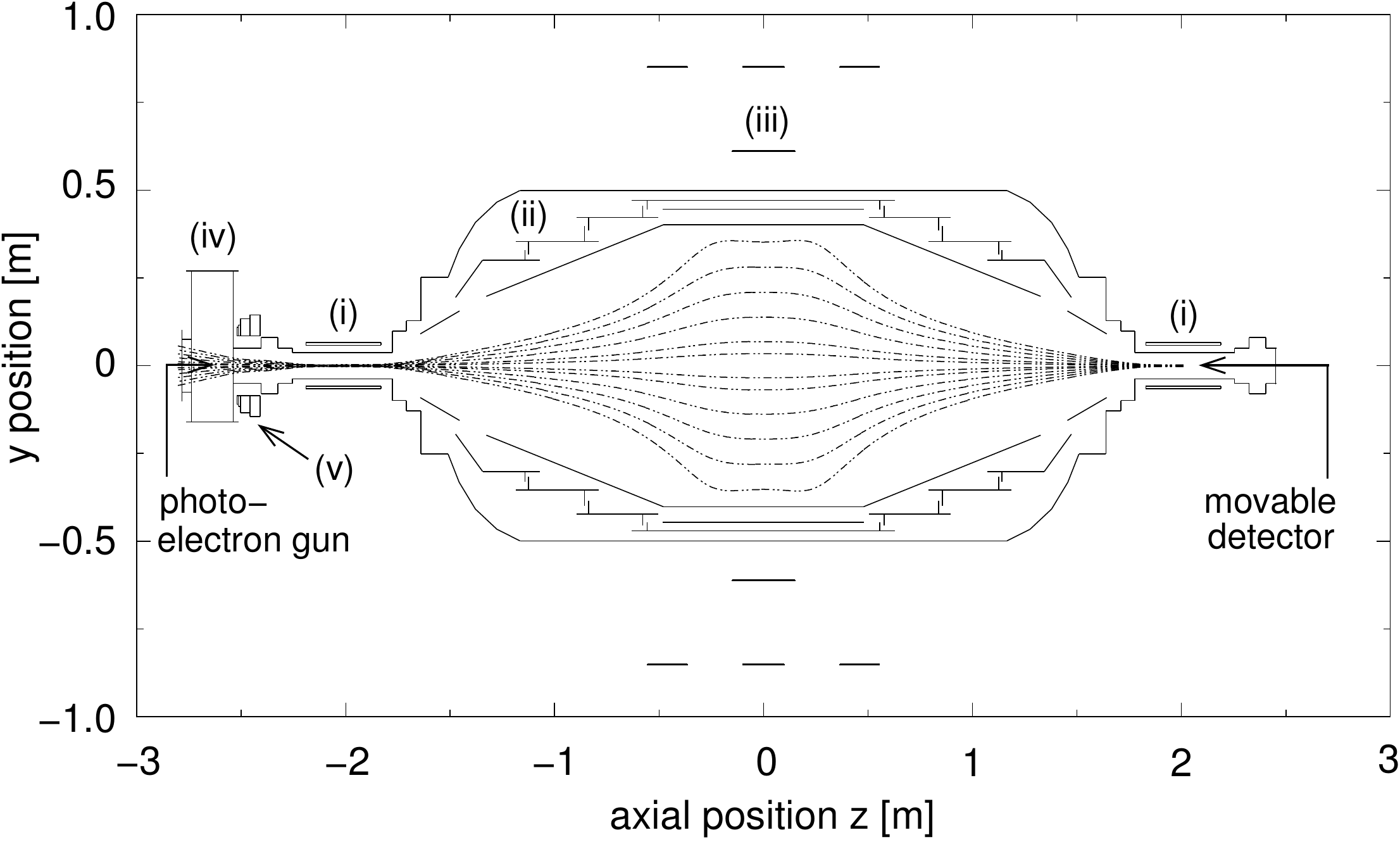}
\caption[Set-up used for the test measurements with the fibre-coupled photoelectron gun at the Mainz {\Mac}.]{Set-up used for the test measurements with the fibre-coupled photoelectron gun at the Mainz {\Mac}. The photoelectron gun and the electron detector were placed at opposite ends of the spectrometer set-up consisting of (i) two superconducting solenoids at the same magnetic field strength, (ii) the spectrometer vacuum tank at ground potential encompassing solid and wire inner electrode systems on high voltage, (iii) a set of field-shaping air coils, (iv) an additional vacuum
chamber and (v) a water-cooled auxiliary coil with current $I_\mathrm{aux.}$ for local enhancement of the magnetic field. The dashed curves indicate magnetic field lines for coil current settings corresponding to a resolving power of $E/\Delta E \approx 2\cdot 10^4$.\label{fig:geometry-mainz}}
\end{figure}

Test measurements to characterise the performance of the prototype electron source were carried out at the \Mac\ of the former Mainz neutrino mass experiment. The experimental set-up for the test measurements is shown in Fig.~\ref{fig:geometry-mainz}. From left to right, the figure includes our fibre-coupled photoelectron source which was inserted into a separate vacuum chamber, the spectrometer, and the electron detector. The polished tips of the optical fibres were coated with a thin film of copper ($\sim$ 20 -- 30\;$\mu$g/cm$^2$) and the LED T9B26C with $\lambda_\mathrm{central} = \unit[265]{nm}$ was used. The current of the auxiliary magnet (cp. Fig.~\ref{fig:geometry-mainz}) was set to $I_\mathrm{aux.} = \unit[100]{A}$ in order to achieve a local enhancement of the magnetic field strength at the location of the electron source. The energy resolution of the electron detector (a Si-PIN diode of the type Hamamatsu S3590-06) was estimated as $\sim 3\;\mathrm{keV}$ (FWHM) at electron energies around $18\;\mathrm{keV}$. This is sufficient to perform a background/noise rejection, but it is not relevant for the determination of the electron spectrum, as the energy measurement is done by the \Mac.

\begin{figure}[!htb]
 \centering
\includegraphics[height= 0.8\textwidth,angle=-90]{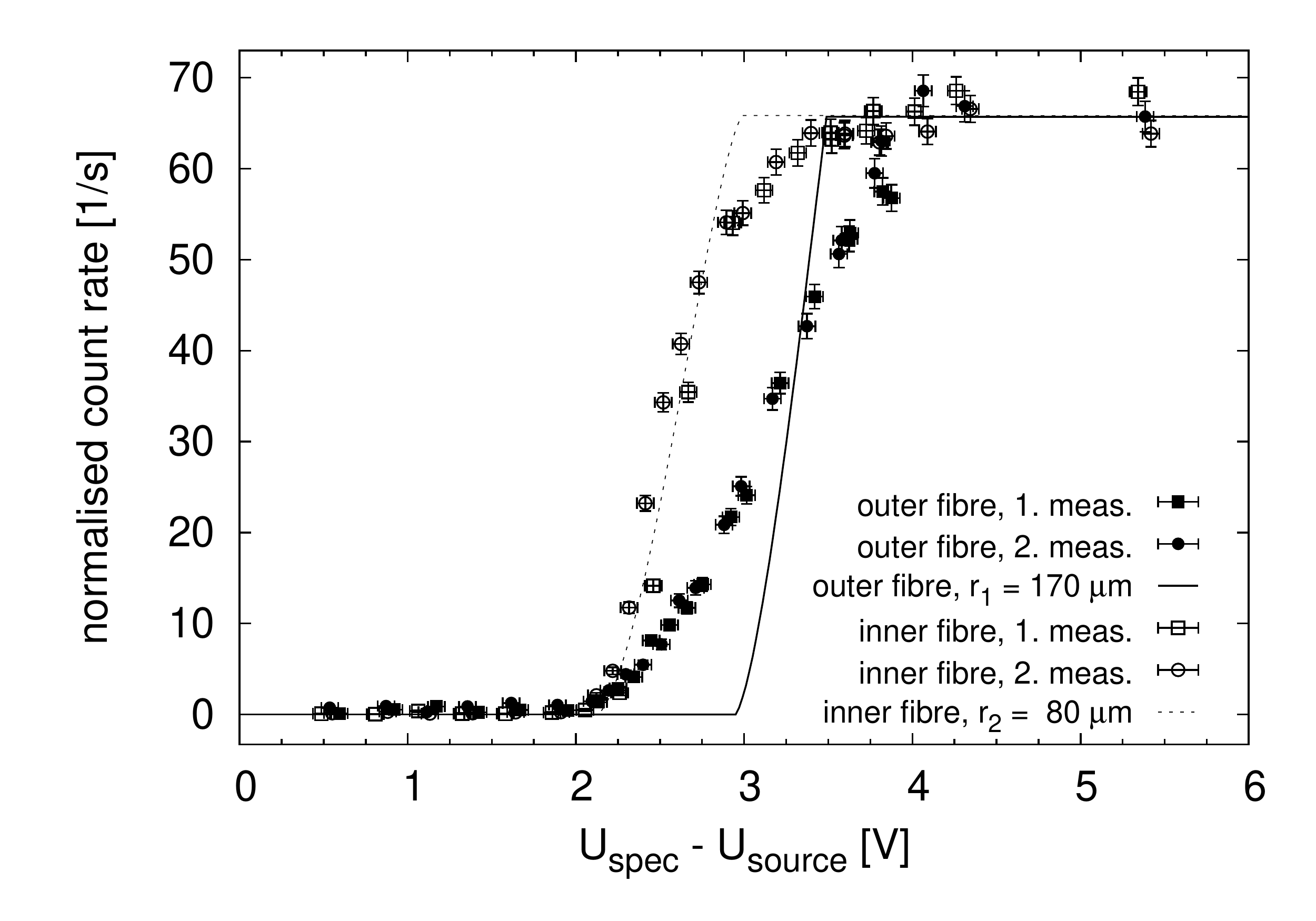}
\caption{Integrated energy spectrum of photoelectrons from the two fibres. The difference of the two (negative) high-voltages $u := U_\mathrm{spec} - U_\mathrm{source}$ representing the abscissa of the plot is a measure of the excess energy $e u$ of the electrons above the filter threshold. Two scans of the spectrum for each fibre are shown. Filled symbols are used for the outer fibre and open symbols for the inner fibre. The measured data are compared with the expected transmission functions, which are calculated for monoenergetic photoelectrons emitted with an angular interval 
corresponding to a central fibre position of $r_1 = 80\;\mathrm{\mu m}$ for the inner fibre (dashed line) and $r_2 = 170\;\mathrm{\mu m}$ for the outer fibre (solid line).\label{fig:fiber-spectra-theotransm}}
\end{figure}

\begin{figure}[!htb]
\centering
\includegraphics[height=0.8\textwidth,angle=-90]{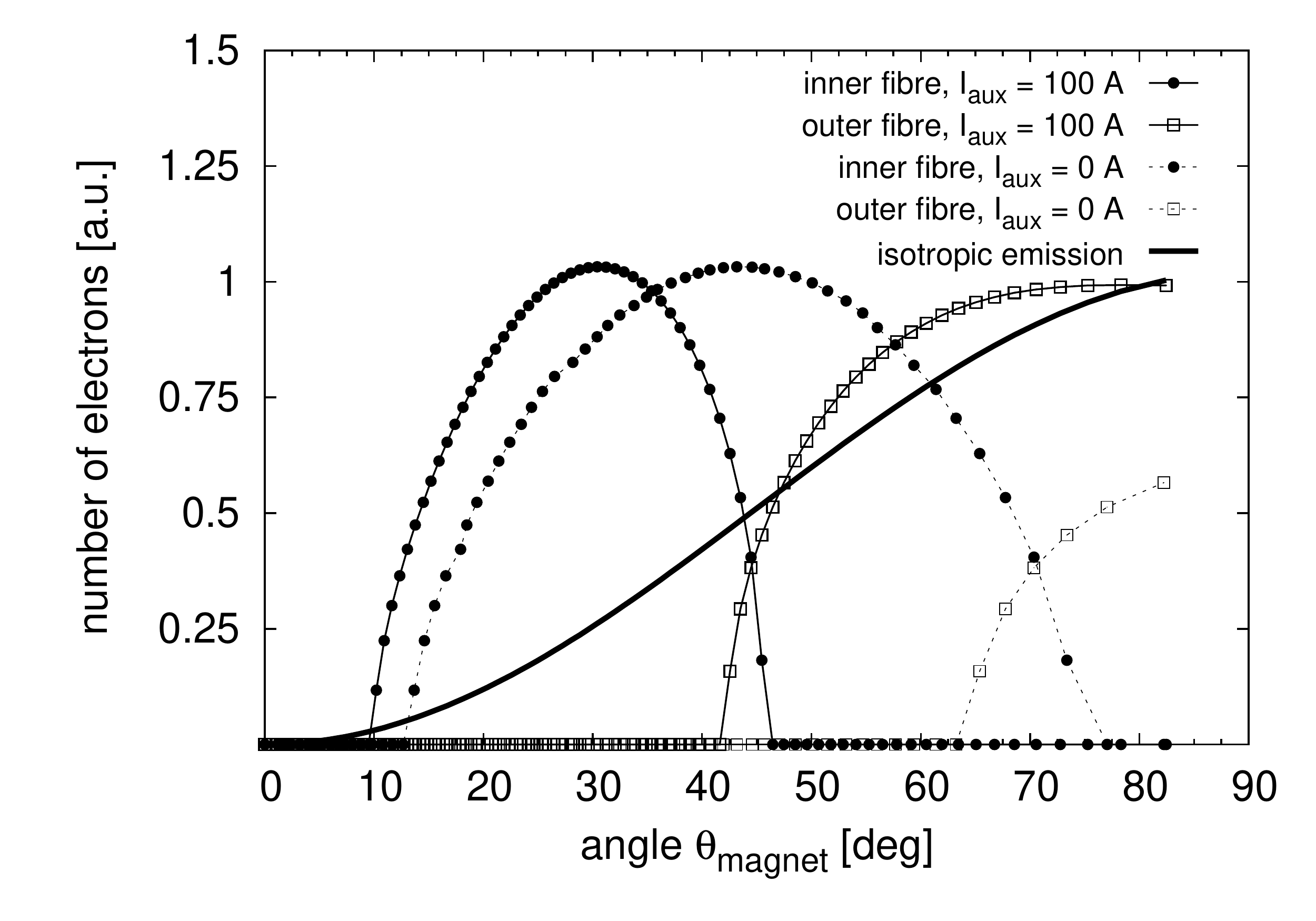}
\caption{Simulated angular distributions at the position of the superconducting
solenoid at the entrance of the Mainz \Mac\
for the two fibres at different settings of the magnetic field (determined by the current $I_\mathrm{aux.}$ 
through the auxiliary coil). For comparison, the expected distribution for an isotropically emitting source is also shown.\label{fig:angular-distr-fibres}}
\end{figure}

By setting the photoelectron source to a fixed high voltage (typically about $-18\;\mathrm{keV}$) and scanning the electrostatic retardation potential of the \Mac\ in small steps around it, we measured several integrated energy spectra for the two intact fibres of the source. The results presented in Fig.~\ref{fig:fiber-spectra-theotransm} clearly show two distinct shapes of the energy spectrum for the two different fibres, which is expected because of their different range of angular emission.
As explained in Section \ref{subsec:mac-e-technique}, the range of starting angles should influence both the onset of the transmission and the width of the transmission curve. Both effects can be observed in the measured data. Transmission of electrons from the outer fibre (\#\,1 in the numbering of Fig.~\ref{fig:drawing-fiberholes}) starts at significantly larger difference
between the electric potential at the photoelectron source and 
at the spectrometer and thus at a significantly higher excess energy above the filter threshold setting than the transmission of electrons from the inner fibre  (\#\,2). This indicates that indeed the starting angles of the electrons from the outer fibre are higher than those of the electrons emitted at the inner fibre. To further investigate this interpretation, we used  computer simulations implementing the detailed geometry of the tip as well as the special field configuration of electric and magnetic fields at the set-up in Mainz in order to estimate the angular emission profile of the individual fibres (see Ref.~\cite{hein09}). The results of these simulations for two different configurations of the magnetic field at the location of the electron souce (normal vs. enhanced field strength) are presented in Fig.~\ref{fig:angular-distr-fibres} and summarised in Table~\ref{tab:angle-values}. The angular emission intervals are rather broad, and they partly overlap. For a weaker magnetic field at the photoelectron source created by setting the auxiliary coil current $I_\mathrm{aux.}$ to zero a large fraction of the photoelectrons are lost due to magnetic reflection, as their angle at the pinch magnet reaches values of $\theta_\mathrm{magnet} = 90^\circ$. 

The lines included in Fig.~\ref{fig:fiber-spectra-theotransm} represent the theoretical transmission functions calculated under the assumption that the photoelectrons are generated with a range of starting angles determined from the simulations. Even though the general trend of the data for the two fibres is described by the theoretical curves, one additional ingredient is obviously still missing in order to make the theoretical expectations match the experimental results: so far, the energy spread of the photoelectrons at their creation was neglected. A much better agreement is achieved by convolving the theoretical transmission curves with a Gaussian distribution of width $\sigma_\mathrm{energy}$, which may take different values for the two individual fibres.

\begin{figure}[!htb]
\centering
\includegraphics[height= 0.8\textwidth, angle=-90]{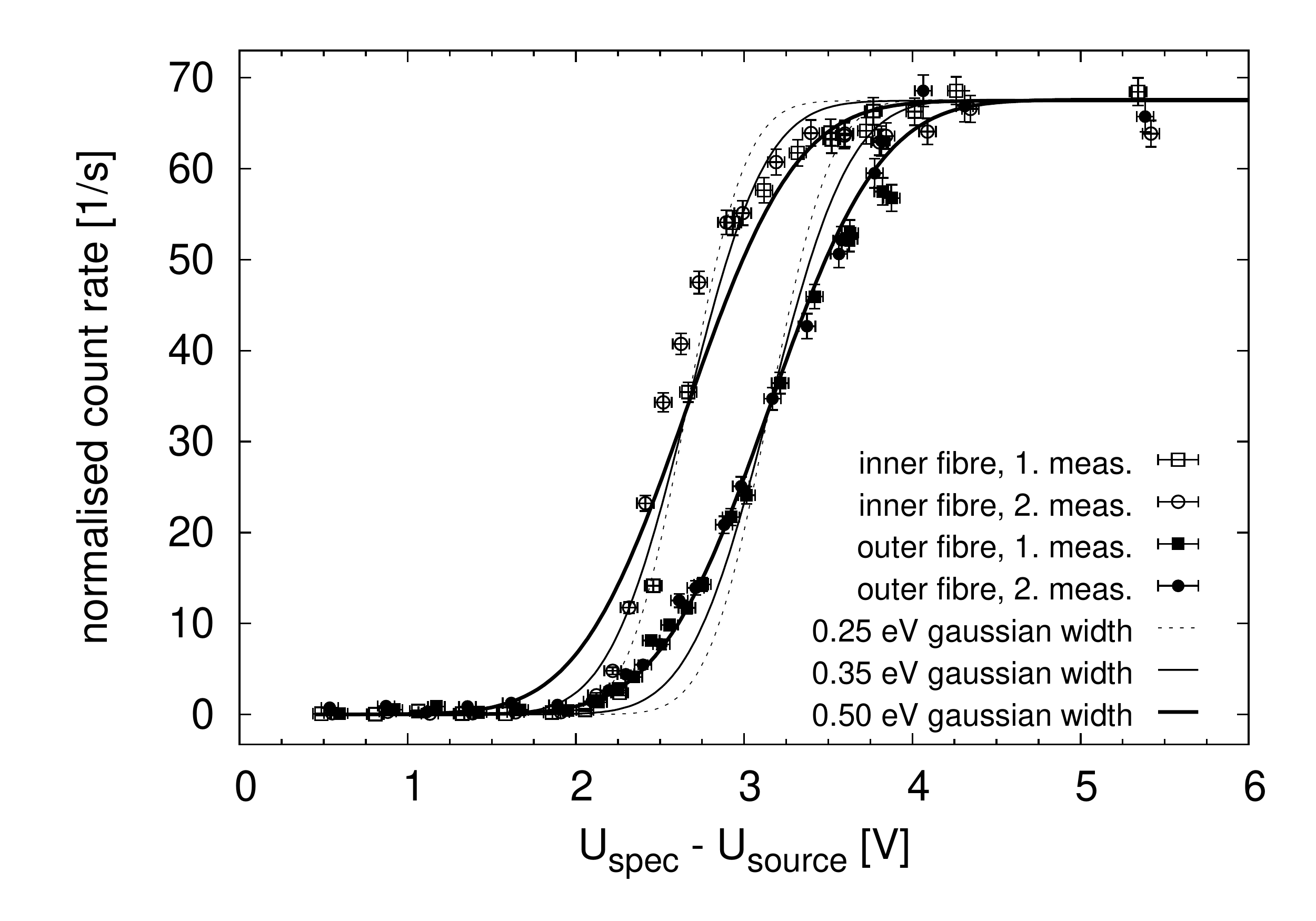}
\caption{Comparison of the measured integrated energy spectra for the photoelectrons from the two fibres with calculated transmission functions smeared out by a Gaussian distribution with varying width $\sigma = 0.25\;\mathrm{eV}$ (dashed lines), $0.35\;\mathrm{eV}$ (thin solid lines) and $0.5\;\mathrm{eV}$ (thick solid lines).\label{fig:fiber-spectra}} 
\end{figure}

In Fig.~\ref{fig:fiber-spectra} the effects of a finite starting energy distribution of the photoelectrons are examined. The same measured data as in Fig.~\ref{fig:fiber-spectra-theotransm} are shown, together with the Gaussian-smeared theoretical transmission curves. By varying the width of the Gaussian to match the measured integrated spectra for the two fibres separately, the energy spread of the photoelectrons can be estimated. Assuming the angular ranges for the two fibres given in Tab.~\ref{tab:angle-values} for the auxiliary coil current of $I_\mathrm{aux.} = 100\;\mathrm{A}$, the values of the energy spreads are derived as $\sigma_\mathrm{energy,\, 1} \approx 0.35$ -- $\unit[0.5]{eV}$ (outer fibre) and $\sigma_\mathrm{energy,\, 2} \approx0.25$ -- $\unit[0.35]{eV}$ (inner fibre), respectively (see \cite{hein09} for details on the simulations). After including the energy spread, the simulated transmission curves match the measured data reasonably well, in particular when considering the significant systematic uncertainties introduced, \eg, by a slight misalignment of the electron source tip with respect to the symmetry axis, deviations of the metal-coated tip from an ideal (perfectly round and smooth) surface, misplacements of the fiber positions on the tip, and the large uncertainty of the magnetic field measurement at the location of the source.

\begin{table}[!htb]
\centering
\caption{Overview of simulation results: angular range of photoelectrons at the position of the spectrometer entrance solenoid. Different magnetic field strengths $B_\mathrm{source}$ were used for the simulations. (Note that $B_\mathrm{source}$ is determined by the setting of the spectrometer solenoid, $B_\mathrm{solenoid} = 6\;\mathrm{T}$, which was left unchanged during the measurements, and by the variable current settings of the auxiliary coil, $0\; \mathrm{A}\leq I_\mathrm{aux.} \leq 120\;\mathrm{A}$.)\label{tab:angle-values}}
\vspace{2mm}
\begin{tabular}{|cc|c|c|}
\hline
$I_\mathrm{aux.}$ [A]	& $B_\mathrm{source}$ [mT] & 	$\theta_\mathrm{magnet,\,1}$ (outer fibre)	& $\theta_\mathrm{magnet,\,2}$ (inner fibre) \\
\hline \hline
$100$ 	& $31 \pm 3$ & $13^\circ - 77^\circ$ & $10^\circ - 46^\circ$\\
$0$ 	& $24 \pm 3$ & $\geq 63^\circ$	& $\geq 42^\circ$\\
\hline
\end{tabular}
\end{table}

\section{Discussion and outlook}
\label{sec:discussion}
In this work we have presented a novel concept of an angular-selective electron source with a narrow energy spread. Test measurements with a prototype source described in this article showed that the idea of selecting specific ranges of transversal kinetic energies by applying non-parallel electric and magnetic fields works. The pickup of transversal energy results from a rapid and non-adiabatic acceleration and is controlled by the strength of the electric field component perpendicular to the magnetic field. We were able to model the angular-dependent emission of electrons and to characterise the properties of the photoelectron source with the help of detailed computer simulations. By comparing the measured electron spectra with simulated ones for varying energy spread and angular distributions, we deduced an intrinsic energy spread of $\delta E \approx \unit[0.35]{eV}$ and clearly distinguished angular emission ranges for the two electron-emitting fibres of the source. At the electron energies around $E = \unit[18]{keV}$ used in our experiments, the measured residual energy spread corresponds to a relative broadening of only $\delta E/E \approx 2 \cdot 10^{-5}$, a benchmark which is very hard to achieve with other electron source concepts such as those based on atomic and/or nuclear standards \cite{venos10,dragoun10}.

The experience gained in the course of the measurements discussed here as well as the detailed simulations allowed us to improve the properties of the electron source. Based on this, photoelectron sources with refined angular selectivity have been developed and tested at the Mainz {\Mac} \cite{hein09,hugenberg-erice09,plate-egun}. 

Such angular-selective electron sources, in particular when combined with the additional feature of pulsed electron emission to allow time-of-flight studies, will become valuable tools for the characterisation of the properties of electron spectrometers (such as the main spectrometer of the KATRIN experiment), but may also prove useful for other experiments.


\acknowledgments
This work was supported by the German Federal Ministry of Education and Research under grant number 05CK5MA/0. We wish to thank the members of AG Quantum/Institut f\"ur Physik, Johannes Gutenberg-Universit\"at Mainz, for their kind hospitality and for giving us the opportunity to carry out these measurements in their laboratory.

\end{document}